\documentclass[aps,prb,twocolumn,showpacs,groupedaddress, amsmath]{revtex4}

\usepackage{graphicx}
\usepackage{amssymb}

\begin{document}

% number in the upper righthand corner of the title page in preprint 
% mode.
%\preprint{}

\title{Aging dynamics of ferromagnetic and reentrant spin glass phases in stage-2 Cu$_{0.80}$C$_{0.20}$Cl$_{2}$ graphite intercalation compound}


\author{Masatsugu Suzuki}
\email[]{suzuki@binghamton.edu}
%\thanks{}
\affiliation{Department of Physics, State University of New York at Binghamton, Binghamton, New York 13902-6000}

\author{Itsuko S. Suzuki}
\email[]{itsuko@binghamton.edu}
%\thanks{}
\affiliation{Department of Physics, State University of New York at Binghamton, Binghamton, New York 13902-6000}


\date{\today}

\begin{abstract}
Aging dynamics of a reentrant ferromagnet stage-2 Cu$_{0.8}$Co$_{0.2}$Cl$_{2}$ graphite intercalation compound has been studied using DC magnetic susceptibility. This compound undergoes successive transitions at the transition temperatures $T_{c}$ ($\approx 8.7$ K) and $T_{RSG}$ ($\approx 3.3$ K). The relaxation rate $S_{ZFC}(t)$ exhibits a characteristic peak at $t_{cr}$ below $T_{c}$. The peak time $t_{cr}$ as a function of temperature $T$ shows a local maximum around 5.5 K, reflecting a frustrated nature of the ferromagnetic phase. It drastically increases with decreasing temperature below $T_{RSG}$. The spin configuration imprinted at the stop and wait process at a stop temperature $T_{s}$ ($<T_{c}$) during the field-cooled aging protocol, becomes frozen on further cooling. On reheating, the memory of the aging at $T_{s}$ is retrieved as an anomaly of the thermoremnant magnetization at $T_{s}$. These results indicate the occurrence of the aging phenomena in the ferromagnetic phase ($T_{RSG}<T<T_{c}$) as well as in the reentrant spin glass phase ($T<T_{RSG}$).
\end{abstract}

%\keywords{}

\maketitle

%\section{}
%\subsection{}
%\subsubsection{}

% If in two-column mode, this environment will change to single-column
% format so that long equations can be displayed. Use
%\begin{widetext}
%\end{widetext}

\section{\label{intro}INTRODUCTION}
Recently the slow dynamics of reentrant ferromagnets has been extensively studied from the time evolution of the magnetization $M(t)$ and the absorption of AC magnetic susceptibility $\chi^{\prime\prime}(t)$ after the appropriate aging protocol.\cite{Jonason1996a,Jonason1996b,Jonason1998,Jonason1999,Vincent2000a,Vincent2000b,Dupuis2002,Suzuki2004,Suzuki2005} The reentrant ferromagnet undergoes two phase transitions at the critical temperatures $T_{RSG}$ and $T_{c}$ ($T_{c}>T_{RSG}$): the reentrant spin glass (RSG) phase below $T_{RSG}$ and the ferromagnetic (FM) phase between $T_{RSG}$ and $T_{c}$. The aging behavior in the RSG phase has been reported in many reentrant ferromagnets including (Fe$_{0.20}$Ni$_{0.80}$)$_{75}$P$_{16}$B$_{6}$Al$_{3}$,\cite{Jonason1996a,Jonason1996b,Jonason1998,Jonason1999} CdCr$_{2x}$In$_{2(1-x)}$S$_{4}$ ($x$ = 0.90, 0.95, and 1.00),\cite{Vincent2000a,Vincent2000b,Dupuis2002} and Cu$_{0.2}$Co$_{0.8}$Cl$_{2}$-FeCl$_{3}$ graphite bi-intercalation compound (GBIC).\cite{Suzuki2004,Suzuki2005} The aging behavior of the RSG phase is similar to that of the spin glass (SG) phase of spin glass systems. In contrast, there have been few reports on the observation of the aging behavior in the FM phase. The measurement of the relaxation rate $S(t)$ [$=(1/H)$d$M(t)$/d$\ln t$] for (Fe$_{0.20}$Ni$_{0.80}$)$_{75}$P$_{16}$B$_{6}$Al$_{3}$,\cite{Jonason1996a,Jonason1996b,Jonason1998,Jonason1999} and Cu$_{0.2}$Co$_{0.8}$Cl$_{2}$-FeCl$_{3}$ GBIC\cite{Suzuki2004,Suzuki2005} has revealed that not only the RSG phase but also the FM phase exhibit aging phenomena. For CdCr$_{2x}$In$_{2(1-x)}$S$_{4}$ with $x$ = 0.90, 0.95, and 1.0,\cite{Vincent2000a,Vincent2000b,Dupuis2002} the aging behavior of the absorption $\chi^{\prime\prime}(\omega,t)$ is observed both in the FM and RSG phases.

In this paper we report our experimental results on the nonequilibrium aging dynamics of the FM phase and the RSG phase in reentrant ferromagnet, stage-2 Cu$_{0.8}$Co$_{0.2}$Cl$_{2}$ graphite intercalation compound (GIC) using DC magnetization measurements. This system undergoes magnetic phase transitions at $T_{RSG} \approx 3.30$ K and $T_{c} \approx 8.70$ K.\cite{Suzuki1998,Suzuki1999} We examine the aging behavior of this system using the zero-field-cooled (ZFC) magnetization measurements. The relaxation rate $S_{ZFC}(t)$ exhibits a peak at a characteristic time $t_{cr}$ below $T_{c}$, indicating the occurrence of the aging phenomena both in the RSG phase and the FM phase. We will also show that the $t$ dependence of $S_{ZFC}(t)$ around $t \gtrsim t_{w}$ is well described by a stretched exponential relaxation, $S_{ZFC}(t) \approx (t/\tau)^{1-n}\exp[-(t/\tau)^{1-n}]$, where $n$ is a stretched exponential exponent and $\tau$ is a relaxation time which is assumed to be equal to $t_{cr}$. The temperature dependence of $t_{cr}$ and $\tau$ shows a very characteristic behavior, which is very similar to that observed in Cu$_{0.2}$Co$_{0.8}$Cl$_{2}$-FeCl$_{3}$ graphite bi-intercalation compound (GBIC).\cite{Suzuki2005} We will also report two kinds of the memory phenomena in the measurement of thermoremnant magnetization (TRM) and the field-cooled (FC) magnetization. When the system is cooled down, a memory of the cooling process is imprinted in the spin structure. This memory is recalled in a continuous heating measurement. The TRM and FC magnetization curves are recovered on heating the system after the specific cooling protocols. The comparison of these curves yields information on the aging and memory effects.

\section{\label{exp}EXPERIMENTAL PROCEDURE}
We used the same sample of stage-2 Cu$_{0.8}$Co$_{0.2}$Cl$_{2}$ GIC that was used in the previous papers.\cite{Suzuki1998,Suzuki1999} The detail of the sample characterization and synthesis were presented there. The stoichiometry of this compound is described by C$_{x}$Cu$_{0.80}$Co$_{0.20}$Cl$_{2}$ with $x=11.04 \pm 0.02$. The $c$-axis repeat distance is equal to $d_{c}=12.83 \pm 0.05 \AA$. The DC magnetization and AC susceptibility were measured using a SQUID magnetometer (Quantum Design, MPMS XL-5) with an ultra low field capability option. The remnant magnetic field was reduced to zero field (exactly less than 3 mOe) at 298 K. The experimental procedure for each measurement is presented in the text and figure captions. The DC magnetic susceptibility for $150\le T\le 298$ K obeys a Curie-Weiss law with the Curie-Weiss temperature and the average effective magnetic moment given by $\Theta =2.52 \pm 0.22$ K and $P_{eff}=2.98 \pm 0.02 \mu_{B}$, respectively. For the AC susceptibility, the amplitude of the AC magnetic field $h$ was 50 mOe and the AC frequency $f$ was ranged between 0.01 and 1000 Hz.

After the ZFC aging protocol (see Sec.~\ref{resultB} for detail), the above conditions, $M_{ZFC}(t)$ was measured as a function of $t$. The relaxation rate defined by $S_{ZFC}(t)$ [$= (1/H)$d$M_{ZFC}(t)$/d$\ln t$] exhibits a peak at a characteristic time $t_{cr}$.\cite{Lundgren1990} Theoretically\cite{Ogielski1985,Koper1988} and experimentally\cite{Chamberlin1984,Hoogerbeets1985a,Hoogerbeets1985b,Lundgren1986,Alba1986,Granberg1987,Alba1987} it has been noticed that the time variation of the ZFC susceptibility $\chi_{ZFC}(t)$ [$=M_{ZFC}(t)/H$] may be described by a stretched exponential relaxation form
\begin{equation} 
\chi_{ZFC}(t) = \chi_{0} - A(t/\tau)^{-m}\exp[-(t/\tau)^{1-n}],
\label{eq01} 
\end{equation} 
where $\chi_{0}$ and $A$ are constants, $m$ may be a positive exponent and is very close to zero, $n$ is a stretched exponential exponent, and $\tau$ is a relaxation time. In the present work, we consider only the case of $m$ = 0 which may be suitable for the description of the long-time aging behavior for $t\gtrsim t_{cr}$. Then the relaxation rate $S_{ZFC}(t)$ can be described by
\begin{equation} 
S_{ZFC}(t)=S_{max}^{0}F(n,t/\tau),
\label{eq01b} 
\end{equation} 
using a scaling function $F(n,\xi)$ [$=e\xi^{1-n}\exp(-\xi^{1-n})$], where $S^{0}_{max}=A(1-n)/e$ and $e$ (= 2.7182) is a basis of natural logarithm. This relaxiation rate $S_{ZFC}(t)$ has a peak ($=S_{max}^{0}$) at a characteristic time $t=\tau$. 

\section{\label{result}RESULT}
\subsection{\label{resultA}Magnetic phase transitions at $T_{RSG}$ and $T_{c}$} 

\begin{figure}
\includegraphics[width=7.0cm]{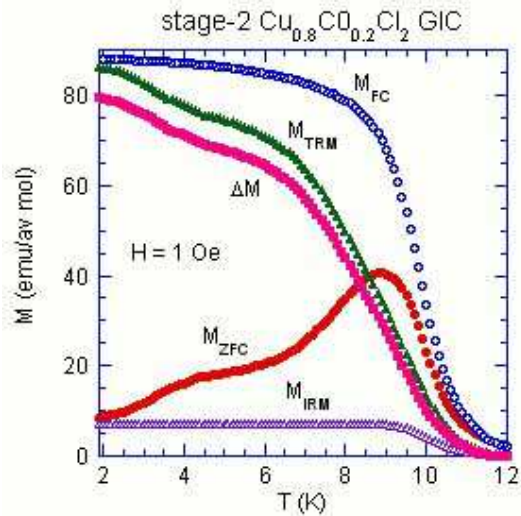}
\caption{\label{fig01}(Color online) $T$ dependence of $M_{FC}$, $M_{ZFC}$, $M_{TRM}$, $M_{IRM}$, and $\Delta M$ ($= M_{FC}-M_{ZFC}$). $H$ (or $H_{c}$) = 1 Oe for stage-2 Cu$_{0.8}$Co$_{0.2}$Cl$_{2}$ GIC. The detail of the cooling protocol for each magnetization is described in the text. The remnant field effect is corrected for each curve of $M$ vs $T$.}
\end{figure}

\begin{figure}
\includegraphics[width=7.0cm]{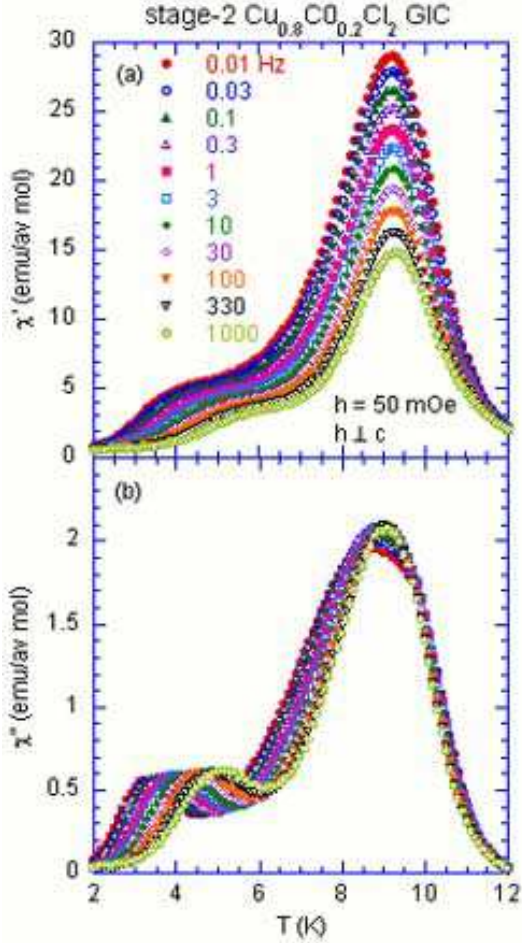}
\caption{\label{fig02}(Color online) $T$ dependence of the AC magnetic susceptibility at various frequencies $f$: $\omega = 2\pi f$. (a) the dispersion $\chi^{\prime}(\omega,T)$ and (b) the absorption $\chi^{\prime\prime}(\omega,T)$. $h$ = 50 mOe. $H$ = 0.}
\end{figure}

We have measured the $T$ dependence of the ZFC magnetization $M_{ZFC}$, the FC magnetization $M_{FC}$, the TRM magnetization $M_{TRM}$, and the isothermal remnant magnetization $M_{IRM}$. We used two types of cooling protocol: (i) the ZFC aging protocol consisting of annealing at 50 K for $1.2\times 10^{3}$ sec and cooling from 50 to 1.9 K in the absence of $H$, and (ii) the FC aging protocol consisting of annealing at 50 K for $1.2\times 10^{3}$ sec and cooling from 50 to 1.9 K in the presence of $H$. The magnetization $M_{ZFC}$ was measured with increasing $T$ in the presence of $H$ (= 1 Oe) after the ZFC aging protocol. The magnetization $M_{TRM}$ was measured with increasing $T$ in the absence of $H$ after the FC aging protocol. The magnetization $M_{FC}$ was measured with decreasing $T$ during the FC aging protocol. The magnetization $M_{IRM}$ was measured with increasing $T$ immediately after the ZFC aging protocol was completed at $T$ = 1.9 K and then the magnetic field $H$ (= 1 Oe) was applied at $T$ = 1.9 K for $t = 1.0 \times 10^{2}$ sec and was turned off. Figure \ref{fig01} shows the $T$ dependence of $M_{ZFC}$, $M_{FC}$, $M_{TRM}$, $M_{IRM}$, and $\Delta M$ ($=M_{FC}-M_{ZFC}$) at $H$ (or $H_{c}$) = 1 Oe. Note that the effect of the remnant magnetic field was corrected by the subtraction of the magnetization $M_{corr}$ which was measured with decreasing $T$ under the remnant magnetic field (3 mOe). The magnetization $M_{ZFC}$ has a shoulder around $T=3.5$ K and a peak at $T$ = 9.0 K. The deviation of $M_{ZFC}$ from $M_{FC}$ appears below $T_{f}$ = 12.5 K, implying that the irreversible effect of magnetization occurs below this temperature. The magnetization $M_{FC}$ drastically increases with decreasing $T$ below 10 K.

Figure \ref{fig02} shows the $T$ dependence of $\chi^{\prime\prime}$. The absorption $\chi^{\prime\prime}$ has two peaks at $T_{RSG}$ and $T_{c}$. The peak at $T_{RSG}$ shifts to the high temperature side with increasing $f$: 3.20 K at $f$ = 0.007 Hz and 5.12 K at 1 kHz. The peak at $T_{c}$ slightly shifts to the high temperature side with increasing $f$: 8.7 K at $f$ = 0.01 Hz and 9.1 K at 1 kHz. The dispersion $\chi^{\prime}$ has a single peak and a shoulder. The peak shifts slightly from 9.20 to 9.30 K with increasing $f$ from 0.01 Hz to 1.0 kHz, while the shoulder shifts greatly from 3 to 6 K. The peak height is strongly dependent on $f$. The detail of the analysis on the $T$ and $f$ dependence of $\chi^{\prime}$ and $\chi^{\prime\prime}$ was described by our previous paper.\cite{Suzuki1998} From these results it may be concluded that our system undergoes phase transitions at $T_{c}$ ($\approx$ 8.7 K) and $T_{RSG}$ ($\approx$ 3.3 K). The low temperature phase below $T_{RSG}$ is a RSG phase and the intermediate phase between $T_{RSG}$ and $T_{c}$ is a FM phase.

\subsection{\label{resultB}Aging behavior of $S_{ZFC}(t)$}
In order to confirm the existence of the aging behavior, we have measured the $t$ dependence of the ZFC magnetization $M_{ZFC}(t)$ at various $T$. Our system was cooled from 50 K to $T$ ($1.9\le T\le 9$ K) in the absence of an external magnetic field. This ZFC cooling process is completed at $t_{a}$ = 0, where $t_{a}$ is defined as an age (the total time after the ZFC aging protocol process). The system is isothermally aged at $T$ until $t_{a}=t_{w}$, where $t_{w}$ ($2.0\times 10^{3}\le t_{w}\le 3.0\times 10^{4}$ sec) is a wait time. The magnetic field $H$ (= 1.0 Oe) is turned on at $t_{a}=t_{w}$ or the observation time $t$ = 0. The ZFC magnetization $M_{ZFC}(t)$ was measured as a function of time $t$. 

\begin{figure}
\includegraphics[width=7.0cm]{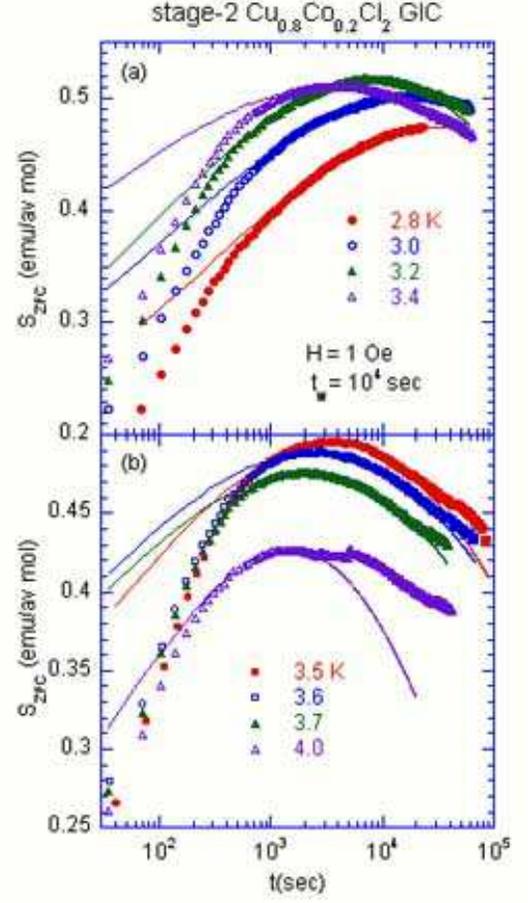}
\caption{\label{fig03}(Color online) $t$ dependence of the relaxation rate $S_{ZFC}$ [$=(1/H)$d$M_{ZFC}(t)$/d$\ln t$] at various $T$. (a) $T$ = 2.8 - 3.4 K. (b) $T$ = 3.5 - 4.0 K. Each measurement was carried out after the ZFC aging protocol: annealing of the system at 50 K for $1.2\times 10^{3}$ sec at $H$ = 0, quenching from 50 K to $T$, and then isothermal aging at $T$ and $H$ = 0 for a wait time $t_{w}$ ($=1.0\times 10^{4}$ sec). The $M_{ZFC}$ measurement was started at $t$ = 0 when the field $H$ (= 1 Oe) is turned on. The solid lines denote the least-squares fitting curves to Eq.(\ref{eq01b}). The parameters $\tau$, $n$ and $S_{max}^{0}$ are given in Fig.~\ref{fig05}.}
\end{figure}

\begin{figure}
\includegraphics[width=7.0cm]{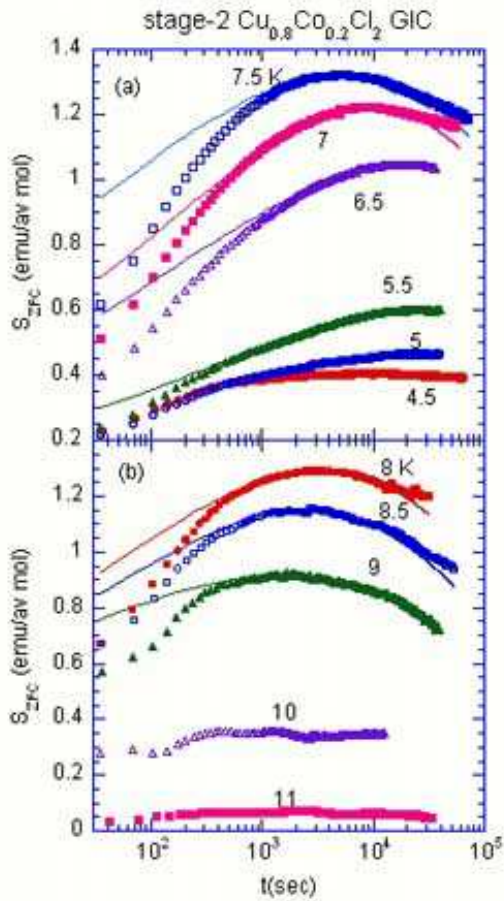}
\caption{\label{fig04}(Color online) $t$ dependence of $S_{ZFC}(t)$ at various $T$. $H$ = 1 Oe. $t_{w}=1.0\times 10^{4}$ sec. (a) $T$ = 4.5 - 7.5 K. (b) $T$ = 8.0 - 11.0 K. The solid lines are curves fitted to Eq.(\ref{eq01b}).}
\end{figure}

\begin{figure}
\includegraphics[width=7.0cm]{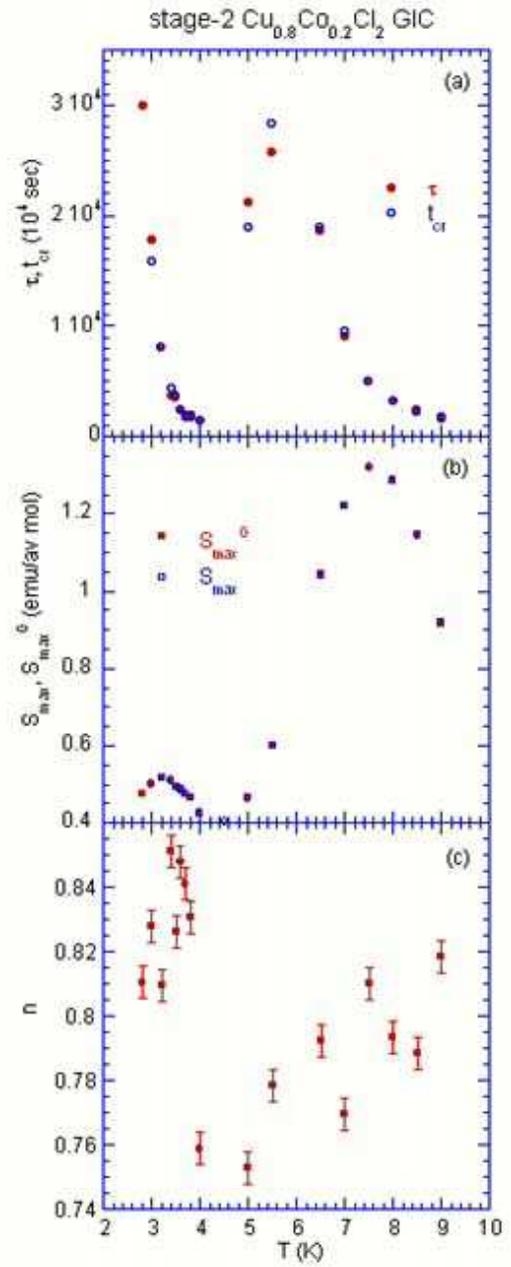}
\caption{\label{fig05}(Color online) $T$ dependence of (a) the peak time $t_{cr}$ ({\Large $\circ$}) and the relaxation time $\tau$ ({\Large $\bullet$}), (b) the maximum value $S_{max}$ ({\Large $\circ$}) and the amplitude $S_{max}^{0}$ ({\Large $\bullet$}), and (c) the stretched exponential exponent $n$. The relaxation rate $S_{ZFC}(t)$ takes a maximum ($S_{max}$) at the peak time $t_{cr}$, where $t_{w}= 1.0\times 10^{4}$ sec and $H = 1$ Oe. The parameters $\tau$, $S_{max}^{0}$, and $n$ are derived from the least-squares fits of the data of $S_{ZFC}(t)$ vs $t$ for $t\gtrsim t_{cr}$ to Eq.(\ref{eq01b}).}
\end{figure}

Figures \ref{fig03} and \ref{fig04} show the $t$ dependence of the relaxation rate $S_{ZFC}(t)$ at various $T$ ($2.8\le T\le 11.0$ K), where $t_{w}=1.0\times 10^{4}$ sec and $H$ = 1 Oe. The relaxation rate $S_{ZFC}(t)$ exhibits a broad peak at a characteristic time $t_{cr}$ in the FM phase as well as in the RSG phase. We find that $S_{ZFC}(t)$ is well described by a stretched exponential relaxation form given by Eq.(\ref{eq01b}) for $t\gtrsim t_{cr}$. Note that the curves of $S_{ZFC}(t)$ vs $t$ greatly deviates from the curves denoted by Eq.(\ref{eq01b}) for $t\ll t_{cr}$. The least-squares fit of these data to Eq.(\ref{eq01b}) yields the parameters $\tau$, $n$ and $S_{max}^{0}$. The solid lines presents the least squares fitting curve to Eq.(\ref{eq01b}). The $T$ dependence of $t_{cr}$, $\tau$, $S_{max}$, $S_{max}^{0}$, and $n$ are shown in Figs.~\ref{fig05}(a), (b), and (c). In Fig.~\ref{fig05}(a) we show the $T$ dependence of $t_{cr}$ and $\tau$ for $t_{w}= 1.0\times 10^{4}$ sec and $H$ = 1 Oe, respectively. The $T$ dependence of $t_{cr}$ is very similar to that of $\tau$ for $2.0\le T\le 9.0$ K. The relaxation time $\tau$ ($\approx t_{cr}$) starts to increase around $T=T_{c}$ with decreasing $T$. It shows a broad peak centered around 5.5 K between $T_{RSG}$ and $T_{c}$, and a local minimum around $T_{RSG}$. It drastically increases with further decreasing $T$ below $T_{RSG}$. Note that very similar behavior of $t_{cr}$ vs $T$ ($\tau$ vs $T$) is also observed in reentrant ferromagnet Cu$_{0.2}$Co$_{0.8}$Cl$_{2}$-FeCl$_{3}$ GBIC.\cite{Suzuki2005} The existence of the broad peak around 5.5 K suggests the chaotic nature of the FM phase in our system (see Sec.~\ref{dis}). The drastic increase of $t_{cr}$ (or $\tau$) below $T_{RSG}$ with decreasing $T$ is a feature common to the SG phases of typical SG systems. Figure \ref{fig05}(b) shows the $T$ dependence of $S_{max}$ (the peak height of $S_{ZFC}(t)$ at $t=t_{cr}$) and $S_{max}^{0}$ for $H$ = 1 Oe and $t_{w}=1.0\times 10^{4}$ sec. We find that the $T$ dependence of $S_{max}^{0}$ agrees well with that of $S_{max}$. The peak height $S_{max}$ at $H$ = 1 Oe exhibits two peaks around $T$ = 3.2 K and at 7.5 K just below $T_{c}$.

In Fig.~\ref{fig05}(c) we show the plot of the exponent $n$ as a function of $T$, where $t_{w}=1.0\times 10^{4}$ sec and $H$ = 1 Oe. The exponent $n$ increases with increasing $T$ and exhibits a peak at $T \approx T_{RSG}$. The exponent $n$ decreases with further increasing $T$. It shows a local minimum around 5.0 K and a local maximum at $T\approx T_{c}$. Similar behavior of $n$ vs $T$ has been reported by Hoogerbeets et al.\cite{Hoogerbeets1985b} for dilute metallic spin glasses. The exponent $n$ increases as $T$ approaches the spin freezing temperature $T_{SG}$ from below.
In summary, we find two relations; $t_{cr}(T) \approx \tau (T)$ and $S_{max}(T) = S_{max}^{0}(T)$. These relations indicate that the stretched exponential relaxation holds well in our system at least for $t\gtrsim t_{cr}$.

\begin{figure}
\includegraphics[width=6.5cm]{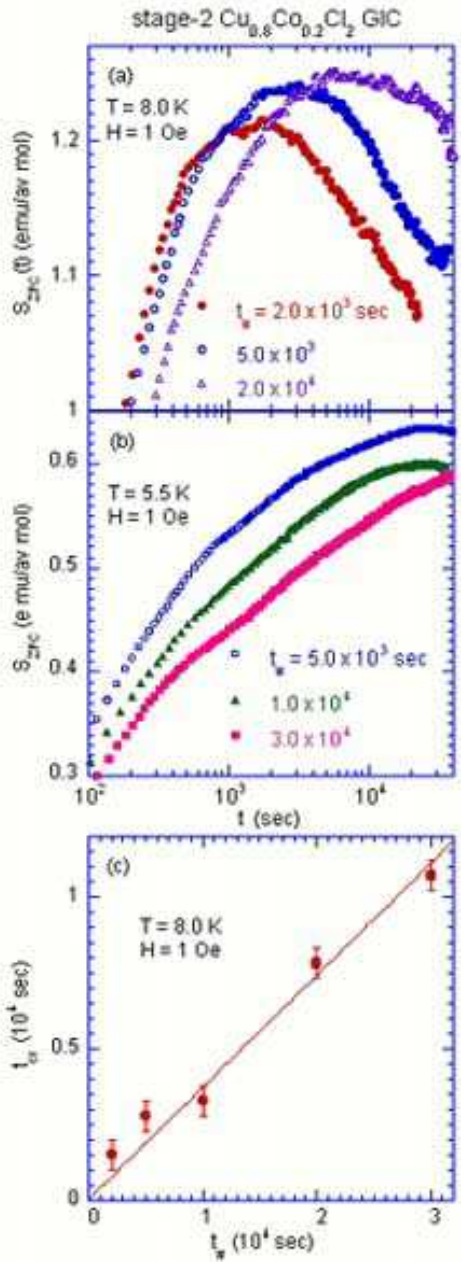}
\caption{\label{fig07}(Color online) $t$ dependence of $S_{ZFC}(t)$ at various wait time $t_{w}$. $H$ = 1 Oe. (a) $T$ = 8.0 and (b) 5.5 K. (c) The peak temperature $t_{cr}$ vs $t_{w}$ at $T$ = 8.0 K, obtained in part from (a).}
\end{figure}

We have measured the $t$ dependence of $M_{ZFC}(t)$ at $T$ = 8.0 and 5.5 K, as the wait time $t_{w}$ is varied as a parameter ($2.0 \times 10^{3}\le t_{w}\le 3.0 \times 10^{4}$ sec). Figures \ref{fig07}(a) and (b) show the $t$ dependence of $S_{ZFC}(t)$ at $T$ = 8.0 and 5.5 K, respectively. As shown in Fig.~\ref{fig07}(a), $S_{ZFC}(t)$ at $T$ = 8.0 K exhibits a peak at $t = t_{cr}$ for each $t_{w}$. This peak shifts to the long-$t$ side with increasing $t_{w}$, showing the aging behavior. In Fig.~\ref{fig07}(c) we show the relation between $t_{cr}$ and $t_{w}$ at $T$ = 8.0 K. The time $t_{cr}$ is proportional to $t_{w}$: $t_{cr}=(0.37 \pm 0.02)t_{w}$. In Fig.~\ref{fig07}(b), in contrast, $S_{ZFC}(t)$ at 5.5 K seems to show a peak at $t = t_{cr} \approx 2.8 \times 10^{4}$ sec for $t_{w} = 1.0 \times 10^{4}$ sec. It is noted that no peak is observed in $S_{ZFC}(t)$ for $t<5.0 \times 10^{4}$ sec for $t_{w} = 3.0 \times 10^{4}$ sec, indicating the divergence of the relaxation time due to the disordered nature of the FM phase.

\subsection{\label{resultC}Genuine TRM measurement}

\begin{figure}
\includegraphics[width=6.5cm]{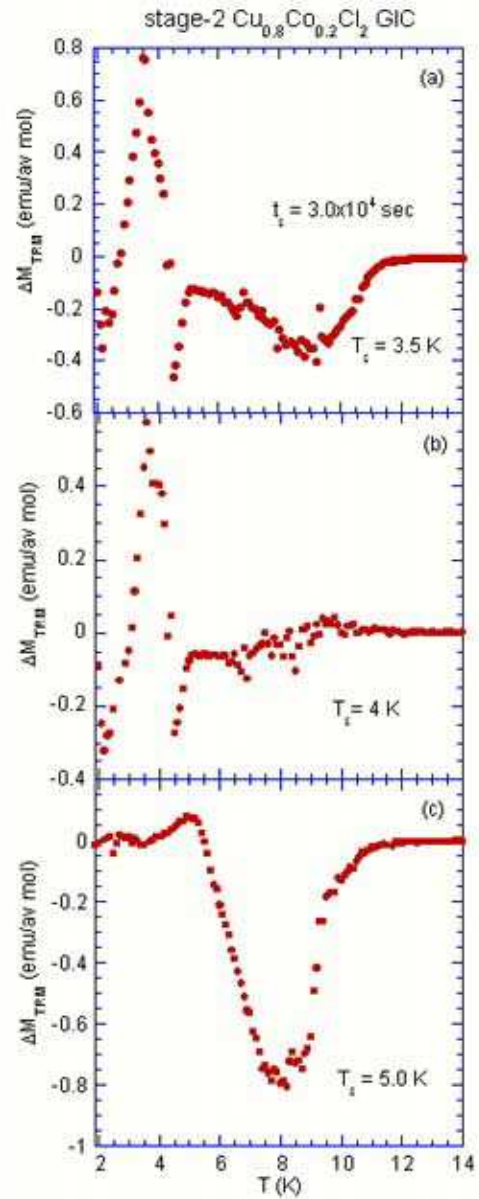}
\caption{\label{fig08}(Color online) $T$ dependence of the difference $\Delta M_{TRM}(T;T_{s},t_{s})$ [$=M_{TRM}(T;T_{s},t_{s})-M_{TRM}^{ref}(T)$]. $t_{w}=3.0\times 10^{4}$ sec. $H_{c}$ = 1 Oe. (a) $T_{s}$ = 3.5 K, (b) $T_{s}$ = 4.0 K, and (c) $T_{s}$ = 5.0 K. $M_{TRM}(T;T_{s},t_{s})$ is measured with increasing $T$ at $H$ = 0 from 2.0 K, after the FC cooling protocol at $H_{c}$ = 1 Oe with a stop-wait at the stop temperature $T_{s}$ for a wait time $t_{s}=3.0\times 10^{4}$ sec. $M_{TRM}^{ref}(T)$ is measured with increasing $T$ at $H$ = 0 from 2.0 K, after the FC aging protocol at $H_{c}$ = 1 Oe without such a stop-wait procedure.}
\end{figure}

\begin{figure*}
\includegraphics[width=12.0cm]{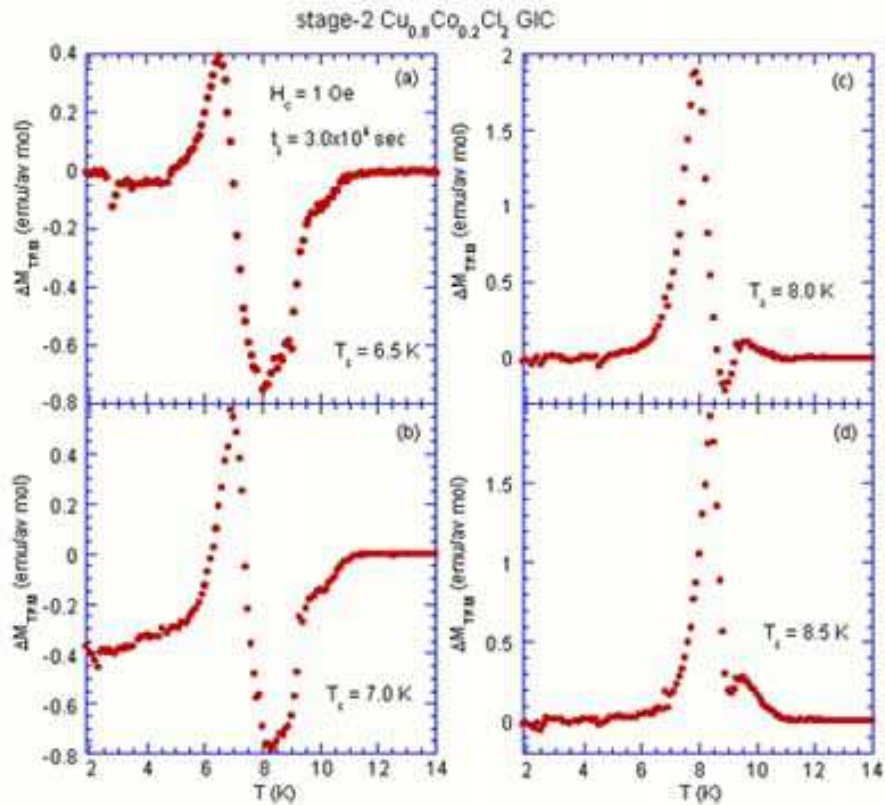}
\caption{\label{fig09}(Color online) $T$ dependence of the difference $\Delta M_{TRM}(T;T_{s},t_{s})$ [$=M_{TRM}(T;T_{s},t_{s})-M_{TRM}^{ref}(T)$]. $t_{s}=3.0\times 10^{4}$ sec. $H_{c}$ = 1 Oe. (a) $T_{s}$ = 6.5 K, (b) $T_{s}$ = 7.0 K, (c) $T_{s}$ = 8.0 K, and (d) $T_{s}$ = 8.5 K. The definition of $M_{TRM}(T;T_{s},t_{s})$ and $M_{TRM}^{ref}(T)$ is the same as for Fig.~\ref{fig08}.}
\end{figure*}

In order to examine the aging and memory effects, we have carried out the genuine TRM measurement. The sample was first rapidly cooled in the presence of $H_{c}$ (= 1.0 Oe) from 50 K. The FC aging protocol was interrupted by stop and wait at an intermittent stop temperature $T_{s}$ ($3.5\le T_{s}\le 8.5$ K). After the isothermal aging for a wait time $t_{s}$ ($=3.0\times 10^{4}$ sec) at $T_{s}$, the cooling of the system was resumed from $T_{s}$ down to 2.0 K. At 2.0 K the magnetic field was switched off. Subsequently the TRM magnetization was measured with increasing $T$ from 2.0 to 14.0 K at $H$ = 0 [$M_{TRM}(T;T_{s},t_{s})$ as the single stop curve]. The result is compared with the TRM magnetization without any intermittent stop during the FC cooling protocol [$M_{TRM}^{ref}(T)$ as the reference curve]. Figures \ref{fig08} and \ref{fig09} present the difference curves $\Delta M_{TRM}(T;T_{s},t_{s})$ with $t_{s}=3.0\times 10^{4}$ sec, obtained by subtracting the reference curve from the single stop curves for different stop temperatures $T_{s}$, where $\Delta M_{TRM}(T;T_{s},t_{s})$ is defined as
\begin{equation} 
\Delta M_{TRM}(T,t_{s})=M_{TRM}(T;T_{s},t_{s})-M_{TRM}^{ref}(T).
\label{eq02} 
\end{equation} 
The differences $\Delta M_{TRM}(T;T_{s},t_{s})$ at $T_{s}$ = 3.5, 4.0, 8.0 and 8.5 K exhibit a positive sharp peak at a temperature close to $T_{s}$. Similar memory effect are observed in the genuine ZFC magnetization measurement for the 3D Ising spin glass Fe$_{0.55}$Mn$_{0.45}$TiO$_{3}$.\cite{Mathieu2001} 

In contrast, the differences $\Delta M_{TRM}(T;T_{s},t_{s})$ at $T_{s}$ = 5.0, 6.5 and 7.0 K are rather different from those at $T_{s}$ = 3.5, 4.0, 8.0, and 8.5 K. They exhibit a relatively broad peak near $T=T_{s}$ as well as a negative local minimum around $T$ = 8 K. The broad peak at $T=T_{s}$ for $T_{s}$ = 5.0, 6.5, and 7.0 K may be closely related to the divergence of $t_{cr}$ and $\tau$ around $T$ = 5.5 - 6.0 K. Such a slow dynamics between $T_{RSG}$ and $T_{c}$ may be related to a possible FM ordered clusters coupled with random dipole-dipole interaction. This FM state smoothly changes into a conventional SG state below $T_{RSG}$. The cause of the local minimum around 8.0 K in $\Delta M_{TRM}(T;T_{s},t_{s})$ is not certain in the present stage.

The ordered domains generated at $T=T_{s}$ are frozen in and survives the spin reconfiguration occurring at lower temperature on shorter length scales. The rejuvenation of the system occurs as the temperature is decreased away from $T_{s}$. The spin configuration imprinted at $T_{s}$ is recovered on reheating. In this sense, the system sustains a memory of an equilibrium state reached after a stop-wait process at $T_{s}$. The influence of the spin configuration imprinted at a stop-wait protocol is limited to a restricted temperature range around $T_{s}$ on reheating. The width of this region may be assigned to the existence of an overlap between the spin configuration attained at $T_{s}$ and the corresponding state at a very neighboring temperature ($T_{s}+\Delta T$). The overlap length $L_{\Delta T}$ is inversely propotional to $\mid\Delta T\mid$.\cite{Lundgren1990} In our system, the spin configuration imprinted during the stop-wait protocol at $T=T_{s}$ for $t=t_{s}$ is unaffected by a small temperature shift such that the overlap length $L_{\Delta T}$ is larger than the average domain sizes. There is a sufficient overlap between the equilibrium spin configurations at the two temperatures $T_{s}$ and $T_{s}+\Delta T$. The situation is different when the temperature shift becomes large. The overlap length becomes shorter than the original domain sizes. A smaller overlap between spin configurations promotes the formation of broken domains. When the temperature shift is sufficiently large, the overlap length is much shorter than the original domain sizes, leading to the rejuvenation of the system.\cite{SuzukiAR}

\subsection{\label{resultD}Memory effect of FC magnetization}

\begin{figure}
\includegraphics[width=7.0cm]{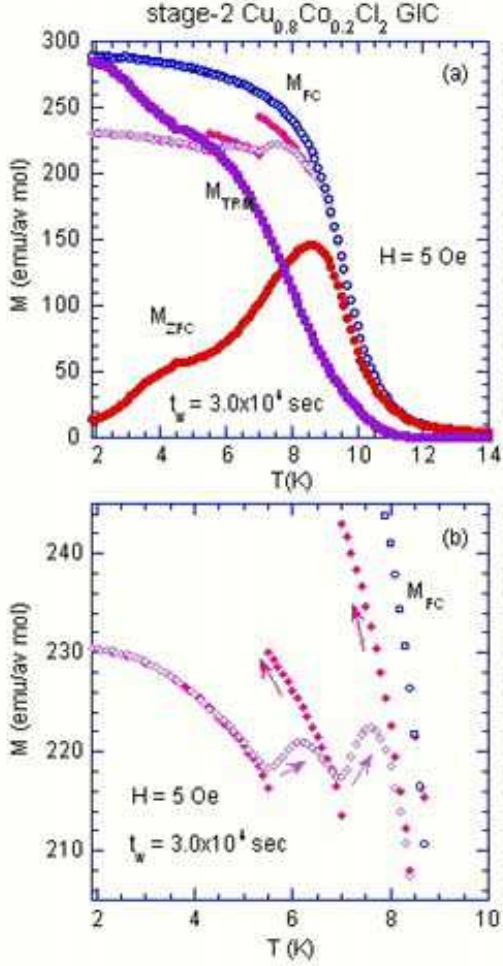}
\caption{\label{fig10}(Color online) (a) and (b) $T$ dependence of $M_{FC}^{IS}(T\downarrow)$ ($\blacklozenge$) and $M_{FC}^{IS}(T$$\uparrow)$ ($\Diamond$) observed in the following FC cooling protocol. The system was quenched from 50 to 15 K in the presence of $H$ (= 5 Oe). $M_{FC}^{IS}(T$$\downarrow)$ was measured with decreasing $T$ from 15 to 1.9 K but with intermittent stops at $T_{s}$= 7.0  and 5.5 K for a stop time $t_{s}=3.0\times 10^{4}$ sec. The field is cut off during each stop. $M_{FC}^{IS}(T$$\uparrow)$ was measured at $H$ = 5 Oe with increasing $T$ after the above FC cooling process. For comparison, the $T$ dependence of $M_{FC}$, $M_{TRM}$ and $M_{ZFC}$ at $H$ (or $H_{c}$) = 5 Oe are also shown as reference curves for comparison. These references curves are measured after either usual FC cooling or usual ZFC protocol without intermittent stop.}
\end{figure}

We have also examined the memory effect of the FC magnetization under the FC aging protocol with a stop and wait process. Our result is shown in Fig.~\ref{fig10}. Our system was cooled through the FC aging protocol from 50 K in the presence of $H$ = 5 Oe. When the system was cooled down to intermittent stop temperatures $T_{s}$ (= 8.5, 7.0 and 5.5 K), the field was cut off ($H$ = 0) and kept at $T$ for $t_{w}$ ($=3.0\times 10^{4}$ sec). In this case, the magnetization $M_{FC}^{IS}(T$$\downarrow)$ decreases with time due to the relaxation. After the wait time $t_{w}$ at each stop temperature, the field ($H$ = 5 Oe) was applied again and the FC cooling process was resumed. Such a FC cooling process leads to a step-like behavior of $M_{FC}^{IS}(T$$\downarrow)$ curve. The value of $M_{FC}^{IS}(T$$\downarrow)$ after resuming below $T$ = 5.5 K behaves almost in parallel to that of the FC magnetization without the intermittent stops ($M_{FC}$ curve as the reference, see Fig.~\ref{fig10}(a)). After reaching the 1.9 K, the magnetization $M_{FC}^{IS}(T$$\uparrow)$ was measured in the presence of $H$ (= 5 Oe) as the temperature is increased at the constant rate (0.05 K/min). The magnetization $M_{FC}^{IS}(T$$\uparrow)$ thus measured exhibits a peak at a characteristic temperature $T_{a}$ = 6.2 K between the stop temperatures $T_{s}$ = 5.5 and 7.0 K, a peak at $T_{a}$ = 7.6 K between $T_{s}$ = 7.0 and 8.5 K, and a kink at $T_{a}$ = 8.8 K above $T_{s}$ = 8.5 K. It is assumed that the anomaly of $M_{FC}^{IS}(T$$\uparrow)$ at $T=T_{a}$ is related to the spin configuration imprinted at $T=T_{s}$ ($<T_{a}$) and $H$ = 0 for a wait time $t_{w}$ during the cooling process. If a temperature difference $\Delta T$ is defined as $\Delta T=T_{a}-T_{s}$, we have $\Delta T$ = 0.7 K for the peak at $T_{a}$ = 6.2 K ($T_{s}$ = 5.5 K), $\Delta T$ = 0.6 K for the peak at $T_{a}$ = 7.6 K ($T_{s}$ = 7.0 K), and $\Delta T$ = 0.3 K a cusp at $T_{a}$ = 8.8 K ($T_{s}$ = 8.5 K). The difference $\Delta T$ tends to decrease with increasing the stop temperature $T_{s}$.

In summary, the spin configuration imprinted at the intermittent stop at $T_{s}$ for a wait time $t_{w}$ at $H$ = 0 during the cooling process strongly affects the $T$ dependence of $M_{FC}(T$$\uparrow)$ when the temperature is increased, exhibiting a peculiar memory effect. Our result is qualitatively in agreement with the results reported by Sun et al.\cite{Sun2003} for superparamagnet (Ni$_{81}$Fe$_{19}$) nanoparticles. Sasaki et al.\cite{Sasaki2005} has proposed a model that the aging and memory effects of such systems may originate solely from a broad distribution of relaxation times in ferromagnetic domains. 

\section{\label{dis}DISCUSSION}
First our results on the aging behavior of $S_{ZFC}(t)$ vs $t$ are compared with those observed in two typical reentrant ferromagnets. The first case is the result of Cu$_{0.2}$Co$_{0.8}$Cl$_{2}$-FeCl$_{3}$ GBIC ($T_{RSG}$ = 3.5 K, $T_{c}$ = 9.7 K).\cite{Suzuki2005} The aging behavior of $S_{ZFC}(t)$ is observed in both RSG phase and the FM phase. The peak time $t_{cr}$ for $t_{w}=3.0 \times 10^{4}$ sec shows a broad peak centered around 4 - 5 K between $T_{RSG}$ and $T_{c}$, and a local minimum around $T_{RSG}$. It decreases with further increasing $T$ below $T_{RSG}$. The broad peak in $t_{cr}$ around 4 - 5 K suggests the chaotic nature of the FM phase. The peak height $S_{max}$ at $H$ = 1 Oe exhibits two peaks around $T=T_{RSG}$ and at 7.0 K just below $T_{c}$, independent of $t_{w}$ ($=1.5\times 10^{4}$ sec or $3.0\times 10^{4}$ sec). The second case is the result of (Fe$_{0.20}$Ni$_{0.80}$)$_{75}$P$_{16}$B$_{6}$Al$_{3}$ ($T_{RSG}$ = 14.7 K and $T_{c}$ = 92 K).\cite{Jonason1996a,Jonason1996b,Jonason1998,Jonason1999} The aging behavior of $S_{ZFC}(t)$ vs $t$ is also observed in both RSG phase and the FM phase. The peak time $t_{cr}$ shows a local minimum around 23 K, and increases with further increasing $T$ between 25 and 30 K. Although no data have been reported for $S_{ZFC}(t)$ vs $t$ above 30 K, it is assumed that $t_{cr}$ shows a local maximum between 30 K and $T_{c}$, since $t_{cr}$ should reduce to zero well above $T_{c}$. The peak height $S_{max}$ ($H$ = 0.5 Oe and $t_{w} = 1.0 \times 10^{3}$ sec) exhibits a peak at 13 K just below $T_{RSG}$, having a local minimum at 25 K, and tends to increase with further increasing $T$. It is assumed that $S_{max}$ shows a local maximum between 30 K and $T_{c}$, since $S_{max}$ should reduce to zero above $T_{c}$. These two peaks of $S_{max}$ vs $T$ are similar to two local maxima around $T_{RSG}$ and between $T_{RSG}$ and $T_{c}$ in our system. 

The features of the aging behavior common to the above two reentrant ferromagnets as well as our system are as follows. The peak time $t_{cr}$ drastically increases with decreasing $T$ below $T_{RSG}$. In this sense, the RSG phase below $T_{RSG}$ is a normal SG phase. The dynamic nature of the FM phase is rather different from that of an ordinary ferromagnet. The peak time $t_{cr}$ as a function of $T$ exhibits a local maximum between $T_{RSG}$ and $T_{c}$. The FM phase just above $T_{RSG}$ shows a dynamic behavior characterized by an aging effect and chaotic nature similar to that of RSG phase.

According to a model proposed by Aeppli et al.,\cite{Aeppli1983} the RSG phase is caused by the random field effect. The FM order is broken down by a random molecular field due to the freezing of spins in the PM clusters which do not contribute to the FM spin order. In the high temperature FM phase the fluctuations of the spins in the PM clusters are so rapid that the FM network is less influenced by them and their effect is only to reduce the net FM moment. On approaching $T_{RSG}$ the thermal fluctuations of the spins in the PM clusters become slower and the coupling between these spins and the FM network becomes significant. Then the molecular field from the slow PM spins acts as a random magnetic field, causing a break up of the FM network into finite domains. 

\section{\label{conc}CONCLUSION}
Stage-2 Cu$_{0.2}$Co$_{0.8}$Cl$_{2}$ GIC undergoes successive transitions at the transition temperatures $T_{c}$ ($\approx$ 8.7 K) and $T_{RSG}$ ($\approx$ 3.3 K). The relaxation rate $S_{ZFC}(t)$ exhibits a characteristic peak at $t_{cr}$ below $T_{c}$, indicating the occurrence of aging phenomena in both the RSG and the FM phases. The relaxation rate $S_{ZFC}(t)$ is well described by a stretch exponential relaxation only for and $t\gtrsim t_{cr}$. The peak time $t_{cr}$ as a function of $T$ exhibits a local maximum around 5.5 K, reflecting a flustrated nature of the FM phase. This result is also supported by the $T$ dependence of the genuine TRM magnetization. It exhibits a sharp peak at $T=T_{s}$ when the stop temperature $T_{s}$ is close to $T_{c}$ and $T_{RSG}$. This peak becomes very broad at $T$ centered around $T_{s}$ ($\approx 5.5$ K). The ordered domains generated at $T=T_{s}$ ($T_{RSG}<T<T_{c}$) are frozen in and survives the spin reconfiguration occurring at lower temperature on shorter length scales. The rejuvenation of the system occurs as the temperature is decreased away from $T_{s}$. The spin configuration imprinted at $T_{s}$ is recovered on reheating, indicating the memory effect.

\begin{acknowledgments}
We would like to thank H. Suematsu for providing us with single crystal kish graphite, and T. Shima and B. Olson for their assistance in sample preparation and x-ray characterization.
\end{acknowledgments}

\end{document}